\documentclass[11pt]{article}

\usepackage{geometry}
\newcommand{\be}{\begin{equation}}
\newcommand{\ee}{\end{equation}}
\usepackage{amsmath}
\usepackage{amssymb}
\usepackage{latexsym}
\usepackage{epsfig}
\usepackage{slashbox}
\usepackage[vcentermath]{youngtab}

\usepackage{slashed}

\usepackage{amsmath,amssymb,amsbsy}
\usepackage{latexsym,layout,amsfonts,amssymb,fontenc,xcolor,amsthm,multirow,amsfonts,bm,textcomp}
\usepackage{hyperref}
\usepackage{empheq}
\usepackage{accents}

\newsavebox{\uuunit}
\sbox{\uuunit}
    {\setlength{\unitlength}{0.825em}
     \begin{picture}(0.6,0.7)
        \thinlines
        \put(0,0){\line(1,0){0.5}}
        \put(0.15,0){\line(0,1){0.7}}
        \put(0.35,0){\line(0,1){0.8}}
       \multiput(0.3,0.8)(-0.04,-0.02){12}{\rule{0.5pt}{0.5pt}}
     \end {picture}}

\def\2{\frac12}
\def\4{\frac14}

\def\equationautorefname~#1\null{eq.~(#1)\null
}

\hypersetup{
    bookmarks=false,         
    unicode=false,          
    pdftoolbar=true,        
    pdfmenubar=true,        
    pdffitwindow=false, 
    pdfstartview={FitH},    
    pdftitle={Wrapping rules (in) string theory},    
    pdfauthor={Eric A.Bergshoeff,  Fabio Riccioni},     
    pdfsubject={},   
    pdfnewwindow=true,      
    colorlinks=false,       
    linkcolor=blue,          
    citecolor=blue,        
    filecolor=blue,      
    urlcolor=blue           
    linkbordercolor={blue},
    citebordercolor={purple},
    urlbordercolor={gray}
}

\begin{document}

\begin{titlepage}
\begin{center}

\hfill UG-17-27\\

\vskip 1.5cm

{\Large \bf
Wrapping rules (in) string theory
}

\vskip 1.5cm

{\bf  Eric A.~Bergshoeff\,$^1$ and Fabio Riccioni\,$^2$}

\vskip 30pt

{\em $^1$ \hskip -.1truecm Centre for Theoretical Physics,
University of Groningen, \\ Nijenborgh 4, 9747 AG Groningen, The
Netherlands \vskip 5pt }

\vskip 15pt

{\em $^2$ \hskip -.1truecm
 INFN Sezione di Roma,   Dipartimento di Fisica, Universit\`a di Roma ``La Sapienza'',\\ Piazzale Aldo Moro 2, 00185 Roma, Italy
 \vskip 5pt }

\vskip 0.5cm

\small{e.a.bergshoeff@rug.nl,
Fabio.Riccioni@roma1.infn.it}

\vskip 1cm

\end{center}

\vskip 0.5cm

\begin{center} {\bf ABSTRACT}\\[3ex]
\end{center}

In this paper we show that the number of all  1/2-BPS branes in string theory compactified on a torus can be derived by universal wrapping rules whose formulation we present. These rules even apply to branes in less than ten dimensions whose ten-dimensional origin is an exotic brane. In that case the wrapping rules contain an additional combinatorial factor that is related to the highest dimension in which the ten-dimensional exotic brane, after compactification, can be realized as a standard brane. We show that the wrapping rules also apply to cases with less supersymmetry. As a specific example, we discuss the compactification of IIA/IIB string theory on $(T^4/{\mathbb{Z}_2}) \times T^n$.

\end{titlepage}

\newpage
\setcounter{page}{1} \tableofcontents


\setcounter{page}{1} \numberwithin{equation}{section}

\section{Introduction}

The presence in string theory of BPS $p$-branes, {\it i.e.}~objects that extend in $p$ spatial directions preserving portions of supersymmetry,
has played a crucial role in establishing non-perturbative duality relations among perturbatively different string models in various dimensions
\cite{Hull:1994ys,Witten:1995ex,Polchinski:1998rq}. From the low-energy viewpoint, branes in $D$ dimensions with more than one transverse direction correspond to supergravity solutions that are charged with respect to the  potentials of the $D$-dimensional supergravity theory or their magnetic duals. In particular, $(D-3)$-branes, also known as `defect branes', are charged under $(D-2)$-form potentials, which are dual to the scalars, and give rise to solutions which are not asymptotically flat. On top of this, string theory also contains BPS states that are domain walls and space-filling branes, charged under $(D-1)$ and $D$-form potentials, which can be introduced in the gauge algebra of supergravity although they carry no degrees of freedom.  For example, the D8-brane in Type IIA string theory and the D9-brane in Type IIB string theory are charged under the RR potentials $C_9$ and $C_{10}$, respectively.

 For maximal theories, all $(D-1)$ and $D$-form potentials have been determined using various methods. In the ten-dimensional case, they were obtained in \cite{Kleinschmidt:2003mf} by suitably decomposing the very-extended Kac-Moody algebra $E_{11}$ \cite{West:2001as} and in \cite{Bergshoeff:2005ac,Bergshoeff:2006qw,Bergshoeff:2010mv} by imposing the closure of the supersymmetry algebra of supergravity. The $E_{11}$ analysis was subsequently extended to any dimension in
\cite{Riccioni:2007au,Bergshoeff:2007qi} to obtain all the supergravity potentials as representations of the corresponding duality symmetry group $G$, in agreement with the tensor hierarchy that one obtains \cite{deWit:2008ta} using the embedding tensor formalism \cite{Nicolai:2000sc}.
In particular,  in \cite{Riccioni:2007au} the representations  of the potentials in the lower-dimensional theories were shown to arise from the dimensional reduction of both standard potentials and mixed-symmetry potentials in ten dimensions. Such mixed-symmetry potentials follow from the decomposition of the $E_{11}$ algebra \cite{Kleinschmidt:2003mf}, and will be crucial for the analysis carried out in this paper.

The classification of 1/2-BPS branes in maximal supersymmetric theories was performed in \cite{Bergshoeff:2010xc,Bergshoeff:2011qk,Bergshoeff:2012ex}  by demanding that a gauge-invariant Wess-Zumino term consistent with worldvolume supersymmetry can be constructed. This analysis shows that the number of $(D-3)$, $(D-2)$ and $(D-1)$-branes is less than the dimensions of the representation of $G$ of the corresponding $(D-2)$, $(D-1)$ and $D$-form potentials, resprectively. This can be understood from a group-theoretic viewpoint by observing that the components of the potentials that couple to branes correspond to the long weights of the representation
\cite{Bergshoeff:2013sxa}, and in the maximal theory only the $(D-2)$, $(D-1)$ and $D$-form potentials belong to representations whose weights have different lengths. This also gives a simple explanation of the fact that the same classification of branes can be obtained by counting  the real roots of the very extended Kac-Moody algebra $E_{11}$ \cite{Kleinschmidt:2011vu}. This result has also a nice explanation in terms of mixed-symmetry potentials. Indeed, the representations of the $(D-2)$, $(D-1)$ and $D$-forms in lower dimensions are those that receive contributions from the dimensional reductions of mixed-symmetry potentials in the ten-dimensional theory \cite{Riccioni:2007au} and in \cite{Bergshoeff:2011zk,Bergshoeff:2011ee,Bergshoeff:2011se,Bergshoeff:2012ex} it was shown that not all the components of these potentials couple to branes. Specifically,  given a  ten-dimensional mixed-symmetry potential $A_{p,q,r,..}$  in a representation such that $p,q,r, ...$ (with $p\ \geq q \geq r ...$) denote the length of each column of the Young Tableau associated to this representation, this corresponds to a brane if some of the  indices $p$  are compactified and contain all the indices $q$, which themselves contain all the indices $r$ and so on. All the other components correspond to shorter weights of representations of $G$ after dimensional reduction. Branes that couple to mixed-symmetry potentials are typically referred to as {\it exotic} in the literature \cite{Elitzur:1997zn,LozanoTellechea:2000mc,deboer}.

The representations of the supergravity potentials in $D$ dimensions can be decomposed in terms of the perturbative $SO(d,d)$ T-duality subgroup that occurs in the embedding
  \begin{equation}
  G \supset \mathbb{R}^+ \times \text{SO}(d,d) \quad ,
  \label{decsodd}
  \end{equation}
where $d=10-D$ and  $\mathbb{R}^+$ is the dilaton shift symmetry. The long weights of a given $SO(d,d)$ representation that occurs in the decomposition are associated with branes whose tensions $T$ all scale in the same way with respect to the dilaton. In particular, in terms of the non-positive integer $\alpha$
giving the scaling of the tension $T \sim (g_s )^\alpha$ with respect to the  string coupling in the string frame, one gets that the ten-dimensional branes are  the fundamental string with $\alpha=0$,  the D-branes with $\alpha=-1$ and  the NS5-brane with  $\alpha =-2$. The Type IIB theory also possesses a 7-brane with $\alpha=-3$ (the S-dual of the D7-brane) and a 9-brane with $\alpha=-4$ (the S-dual of the D9-brane).
In $D$ dimensions, for each $\alpha \geq -3$  the branes belong to   representations of $SO(d,d)$ whose number of long weights
can all be reproduced starting from the $p$-branes of the ten-dimensional theories by means of the following
``wrapping rules'' \cite{Bergshoeff:2011mh,Bergshoeff:2011ee}
\vskip .2truecm
\begin{align}\label{allwrappingrulesinonego}
\alpha=0 \ \ \  : \quad  & \begin{cases} {\rm wrapped}   \ \rightarrow\   \ {\rm doubled}\\ {\rm unwrapped} \  \rightarrow \  {\rm undoubled} \ ,\end{cases} \nonumber\\~\nonumber\\
\alpha=-1 \ :\quad & \begin{cases} {\rm wrapped}   \ \rightarrow\   \ {\rm undoubled}\\ {\rm unwrapped} \  \rightarrow \  {\rm undoubled} \ ,\end{cases} \\ ~\nonumber\\
\alpha=-2 \ :\quad  & \begin{cases} {\rm wrapped}   \ \rightarrow\   \ {\rm undoubled}\\ {\rm unwrapped} \  \rightarrow \  {\rm doubled} \ ,\end{cases} \nonumber\\~\nonumber
\end{align}
\vfill\eject
\begin{align}
\alpha=-3 \ :\quad & \begin{cases} {\rm wrapped}   \ \rightarrow\   \ {\rm doubled}\\ {\rm unwrapped} \  \rightarrow \  {\rm doubled} \ . \end{cases}\nonumber
\end{align}
\vskip .2truecm
\noindent Here wrapped (unwrapped) means that the brane is compactified over a worldvolume (transverse) direction while doubled (undoubled) means that the brane, after compactification, gives rise to two branes (one brane). The nice thing of these wrapping rules is that they give
the number of branes in any dimension knowing the number of branes in one dimension higher. As far as the $\alpha=-4$ branes are concerned,  there is one particular irreducible representation of space-filling branes that  contains the dimensionally reduced 9-brane of IIB, and whose number is determined by  the additional wrapping rule \cite{Bergshoeff:2012ex}
 \begin{equation}
 \alpha=-4 \ :\quad{\rm wrapped}   \ \rightarrow\   \ {\rm doubled} \quad .\label{wrappingrulealpha=-4}
 \end{equation}
 On top of this there are other $\alpha=-4$ branes that are not related, via any compactification, to the 9-brane of Type IIB string theory.
The final outcome is that the number of all the branes belonging to the $SO(d,d)$ representations that contain, upon compactification,  the branes of the ten-dimensional theories
can be simply obtained using the wrapping rules above.

Recently, in \cite{Lombardo:2016swq} a universal T-duality rule for all the string theory potentials, $p$-form as well as mixed-symmetry potentials, that couple to branes was derived.
 The rule can be stated  as follows:
given  an $\alpha=-n$ brane associated to a mixed-symmetry potential such that the $x$ index occurs $p$ times (in $p$ different sets of antisymmetric indices, i.e.~columns of the corresponding Young tableau),  this is mapped by T-duality along $x$ to the brane associated to the potential in which the $x$ index occurs $n-p$ times. Schematically, this can be written as
\begin{equation}
\alpha=-n \ : \qquad \quad \underbrace{x,x,...,x}_p \ \overset{{\rm T}_x}{\longleftrightarrow} \ \underbrace{x,x, ....,x}_{n-p} \quad . \label{allbranesallalphasruleintro}
\end{equation}
Using this rule, one can determine, starting from any brane, the full set of branes that are related to it by chains of T-duality transformations. In the next section we will give several examples of this universal T-duality rule and, furthermore,  we will show that this rule naturally explains the wrapping rules given above.

All the branes with  $\alpha=-4$ different from those satisfying the wrapping rule in eq.~\eqref{wrappingrulealpha=-4}, as well as all those with $\alpha\leq -4$, are not connected by $SO(d,d)$ transformations to any of the branes of the ten-dimensional theory. The fact that the highest dimension in which all these branes appear is lower than ten means that the $SO(d,d)$ representations arise entirely from mixed-symmetry potentials of the ten-dimensional theory. On the other hand, one can still apply the T-duality rule in eq.~\eqref{allbranesallalphasruleintro} to any of the brane components of such mixed-symmetry potentials to connect it to all the other branes in the same representation. As we will see, this implies that a generalisation of the wrapping rules above can be derived. The rule is the following: starting from the highest dimension in which a brane belonging to the specific representation occurs, eq.~\eqref{allbranesallalphasruleintro} determines whether such brane doubles or not upon dimensional reduction. On top of that, there is an additional universal multiplicity due to the mixed-symmetry indices. Taking into account this generalized rule, we manage to reproduce all the numbers of branes of the maximal theory in any dimension.

After discussing the maximal case, we proceed to apply the wrapping rules to the half-maximal supersymmetric theory corresponding to the compactification of Type IIA/IIB string theory on $(T^4/{\mathbb{Z}_2}) \times T^n$. The number of single-brane states of this theory in any dimension was determined in \cite{Pradisi:2014fqa}. In this reference it was shown that the wrapping rules in eqs.~\eqref{allwrappingrulesinonego} and \eqref{wrappingrulealpha=-4} are still valid if one compactifies on tori starting from the six-dimensional theory. We will show that the number of all the branes in this theory can be determined starting from the highest dimensional representative of each chain of branes by observing from eq.~\eqref{allbranesallalphasruleintro} whether the brane doubles or not, and by computing the multiplicity that results from the mixed-symmetry indices of the corresponding potential.

This paper is organised as follows. In section 2 we consider the maximal case and show that the number of all branes results from the generalised wrapping rules discussed above. We then extend in section 3 this analysis to the half-maximal  case and  in particular we consider the IIA and IIB theories compactified on $(T^4/{\mathbb{Z}_2}) \times T^n$. We show that also in this case the number of all branes can be derived by applying the same generalised wrapping rules. Finally, section 4 contains our conclusions.

\section{Wrapping rules for IIA/IIB on $T^n$}

The aim of this section is to show that the universal T-duality rules in eq.~\eqref{allbranesallalphasruleintro} allow to derive generalised wrapping rules for all the branes in IIA/IIB string theory compactified on a torus. These generalised wrapping rules contain, in addition to a doubling/undoubling rule as those given in eqs.~\eqref{allwrappingrulesinonego} and \eqref{wrappingrulealpha=-4}, a universal multiplicity factor.  We first show that the wrapping rules in eqs.~\eqref{allwrappingrulesinonego} and \eqref{wrappingrulealpha=-4} are a natural consequence of eq.~\eqref{allbranesallalphasruleintro}. We then  move on to consider the branes that are not related by T-duality to any brane of the ten-dimensional theory and show that for these branes the generalised wrapping rules apply.

Following \cite{Bergshoeff:2012ex}, we denote the potentials with $\alpha=0,-1,-2,-3,...$ by $B,C,D,E,...$ and so on. In Table \ref{tabletendimbranes} we list the full set of potentials corresponding to the branes that satisfy the wrapping rules in eqs.~\eqref{allwrappingrulesinonego} and \eqref{wrappingrulealpha=-4}. In particular, in the second column we give the potentials of the $D$-dimensional theory as representations of $SO(d,d)$. The $A$ index is a vector index of $SO(d,d)$ while $a$ and $\dot{a}$ are spinor indices of opposite chirality. The indices $A_1 ... A_n$ are always meant to be completely antisymmetrised, and the potential $F_{D, A_1 ...A_d}^+$ in the last row belongs to the self-dual $SO(d,d)$ representation with $d$ antisymmetric indices.  The branes correspond to the long weights of all these representations \cite{Bergshoeff:2013sxa}. In the IIA/IIB column we list the mixed-symmetry potentials of the ten-dimensional theory that give rise after dimensional reduction to all the components of the $D$-dimensional potentials that correspond to branes. In particular, as reviewed in the introduction the branes
correspond to the components of  the  ten-dimensional mixed-symmetry potential $A_{p,q,r,..}$  such that  some of the  indices $p$  are compactified and contain all the indices $q$, which themselves contain all the indices $r$ and so on.

\begin{table}
\begin{center}
\begin{tabular}{|c||c||c|c|}
\hline \rule[-1mm]{0mm}{6mm} $\alpha$ & potential  & IIA & IIB\\
\hline
\hline \rule[-1mm]{0mm}{6mm}
$0$ & $B_{1,A}$ \ \ \ \ \ \ $B_2$ & \multicolumn{2}{c|}{$B_2$ \ \ \ \  $g_{\mu \nu}$}\\
\hline \rule[-1mm]{0mm}{6mm}
$-1$ & $C_{2n+1, a}$ \ \ \ \ $C_{2n ,\dot{a}}$ & $C_{2n+1}$ & $C_{2n}$  \\
\hline \rule[-1mm]{0mm}{6mm}
$-2$ & $D_{D-4} \ \  D_{D-3,A}  \ \  D_{D-2 , A_1 A_2} \  \  D_{D-1, A_1 A_2 A_3} \  \  D_{D , A_1 ...A_4}$ & \multicolumn{2}{c|}{ $D_{6+n,n}$ }\\
\hline \rule[-1mm]{0mm}{6mm}
$-3$ & $ E_{D-2, \dot{a}} \ \ \ \ E_{D-1, A \dot{a}}\ \  \ \ E_{D, A_1 A_2 \dot{a}}$ & $E_{8+n,2m+1,n}$ & $E_{8+n,2m,n}$\\
\hline \rule[-2mm]{0mm}{7mm}
$-4$ & $ F_{D, A_1 ... A_d}^+$ & $F_{10,2n+1,2n+1}$ & $F_{10,2n,2n}$\\
\hline
\end{tabular}
\end{center}
  \caption{\sl The potentials associated to the branes that satisfy the wrapping rules in eqs.~\eqref{allwrappingrulesinonego} and \eqref{wrappingrulealpha=-4} for the different values of $\alpha$. In the second column we list the $D$-dimensional potentials as representations of $SO(d,d)$,  while in the third and fourth we give the corresponding mixed-symmetry potentials for both the IIA and IIB theory. For $\alpha=0$ and $\alpha=-2$ the potentials for IIA and IIB are the same.
  \label{tabletendimbranes}}
\end{table}

One can obtain the number of independent branes in a given dimension by determining the independent brane components of the corresponding mixed-symmetry potentials. We consider as an example the $\alpha=-2$ case. In ten dimensions only the NS5 brane corresponding to $D_6$ is present in both the IIA and IIB theories, while in $D=9$ one has a 4-brane corresponding to  $D_{5\, x}$ and two 5-branes corresponding to $D_6$ and $D_{6  \, x,x}$. Here we have denoted with $x$ the internal direction, and the 4-brane is the wrapped NS5, while the 5-branes are the unwrapped NS5 and the KK-monopole. Generalising this to all the branes with $\alpha=-2$ in all dimensions one finds that the right number is obtained by applying the $\alpha=-2$ wrapping rule in eq.~\eqref{allwrappingrulesinonego}. The same applies for the branes with different values of $\alpha$.

We now show that the T-duality rules of eq.~\eqref{allbranesallalphasruleintro} discovered in \cite{Lombardo:2016swq} naturally explain all the wrapping rules for the  branes that have a ten-dimensional origin.
Again, we focus for simplicity on the $\alpha=-2$ case. In this case, as we read from Table \ref{tabletendimbranes} the mixed-symmetry potentials are $D_{6+n,n}$ and  the T-duality rules state that
\begin{equation}
0 \rightarrow 1,1 \qquad 1 \rightarrow 1 \qquad .
\end{equation}
The first rule means that if the potential has no indices along $x$, after T-duality this is mapped to a potential with $x$ added on both sets of indices, while the second rule means that if the potential has one index along $x$ only in the first set of indices, this is mapped to the same component of the T-dual theory.
Suppose now that we  compactify from 10 to 9  as discussed above.  By T-duality along $x$, $D_{6}$ goes to $D_{6 \, x,x}$ while $D_{5 \, x}$ is fixed. Therefore the brane doubles when it does not wrap. The same applies for all the $\alpha=-2$ branes in any  dimension.

For $\alpha=-3$ one has
\begin{equation}
0 \rightarrow 1,1,1 \qquad 1 \rightarrow 1,1 \quad , \label{Tdualityrules-3}
\end{equation}
and so there are no potentials that are fixed under T-duality. As a consequence, these branes always double exactly as the wrapping rule in
eq.~\eqref{allwrappingrulesinonego} states. Starting  with $E_8$ in IIB, corresponding to the S-dual of the D7-brane, one gets  in nine dimensions $E_{7 \, x}$, which couples to a 6-brane,  and $E_8$, which couples to a 7-brane. Using eq.~\eqref{Tdualityrules-3}, the former
is mapped to $E_{7 \, x,x}$, and the latter is mapped to $E_{8 \, x,x,x}$.
One therefore gets that  from  the IIA perspective the 6-brane and 7-brane with $\alpha=-3$ in nine dimensions arise from the IIA mixed-symmetry potentials $E_{8,1}$ and $E_{9,1,1}$. In this case T-duality does not generate any doubling, but simply gives a  IIA origin in terms of mixed-symmetry potentials of the same nine-dimensional branes. This is consistent with the fact that there are no $\alpha=-3$ branes in Type  IIA string theory.
By further compactifying to 8 dimensions along the coordinate $y$, one gets that the IIB potential $E_8$ gives $E_{6 \, xy}$, $E_{7\, x}$, $E_{7 \, y}$ and $E_8$. We can now perform two T-dualities along $x$ and $y$ remaining in the same IIB theory. The rules in eq.~\eqref{Tdualityrules-3} give
\begin{eqnarray}
& & E_{6 \, xy} \rightarrow E_{6 \, xy,xy}\nonumber \\
& & E_{7 \, x} \rightarrow E_{7 \, xy ,xy ,y} \nonumber \\
& & E_{7 \, y} \rightarrow E_{7 \, xy ,xy ,x} \nonumber \\
& & E_8 \rightarrow E_{8 \, xy , xy ,xy} \quad ,
\end{eqnarray}
resulting in the components of the potentials $E_{8,2}$, $E_{9,2,1}$ and $E_{10,2,2}$. One can see that all the branes double in going from nine to eight dimensions. The same applies in any other dimension.

Finally, we discuss the $\alpha=-4$ branes that satisfy the wrapping rule in eq.~\eqref{wrappingrulealpha=-4}. In this case the T-duality rule is
\begin{equation}
0 \rightarrow 1,1,1,1 \qquad 1 \rightarrow 1,1,1 \qquad 1,1 \rightarrow 1,1 \quad . \label{Tdualityrules-4}
\end{equation}
In IIB one has a space-filling brane corresponding to the potential $F_{10}$. In nine dimensions this gives $F_{9 \, x}$, which by T-duality is mapped to $F_{9 \, x ,x,x}$ corresponding to the $F_{10,1,1}$ mixed-symmetry potential of the IIA theory. Again, in this case there is no doubling, we just obtain a IIA origin of the same nine-dimensional brane as an exotic brane. By further compactifying to $D=8$, one gets $F_{8\, xy}$, which under two T-dualities is mapped to $F_{8 \, xy,xy,xy}$, which is the brane component of the IIB mixed-symmetry potential $F_{10,2,2}$. This doubling continues to occur in any dimension.

\begin{table}
\begin{center}
\begin{tabular}{|c||c||c|c|}
\hline \rule[-1mm]{0mm}{6mm} $\alpha$ & potential  & IIA & IIB\\
\hline
\hline \rule[-1mm]{0mm}{6mm}
$-4$ & $F_{D-1, A_1 ...A_{d-3}}$\ \ $F_{D,A B_1 ...B_{d-3}}$ & \multicolumn{2}{c|}{$F_{9+n,3+m,m,n}$}\\
\hline \rule[-1mm]{0mm}{6mm}
$-4$ & $F_{D-2, A_1 ...A_{d-6}}$ \ \  $F_{D-1, A B_1 ...B_{d-6}}$ & \multicolumn{2}{c|}{$F_{8+n,6+m,m,n}$} \\
\hline \rule[-1mm]{0mm}{6mm}
$-5$ & $ G_{D, A_1 ...A_{d-4}  \dot{a}}$ & $ G_{10, 4+n,2m+1, n}$ & $ G_{10, 4+n,2m, n}$\\
\hline \rule[-2mm]{0mm}{7mm}
$-5$ & $ G_{D-1, A_1 ...A_{d-6}  a} $ \ \  $G_{D, A B_1 ...B_{d-6}a}$ & $ G_{9+p, 6+n,2m, n,p}$ & $ G_{9+p, 6+n,2m+1, n,p}$ \\
\hline \rule[-2mm]{0mm}{7mm}
$-6$ &$D=4: \ H_{4, A_1 ...A_4} \ \ D=3:\ H_{3, A B_1 ...B_5}$ &   \multicolumn{2}{c|}{$H_{10, 6+n, 2+m,m,n}$} \\
\hline \rule[-2mm]{0mm}{7mm}
$-7$ &$D=4: \ I_{4, \dot{a}} \ \ D=3:\ I_{3, A B \dot{a}}$ &  $I_{10,6+n,6+n,2m+1,n,n}$ &  $I_{10,6+n,6+n,2m,n,n}$\\
\hline
\end{tabular}
\end{center}
  \caption{\sl The potentials associated to all the branes of the maximal theory in four dimensions and above that are not related by T-duality to branes that occur in ten dimensions. In the second column we list the $D$-dimensional potentials as representations of $SO(d,d)$, while in the third and fourth column we list the corresponding ten-dimensional mixed-symmetry potentials. For $\alpha=-4$ and $\alpha=-6$ the IIA and IIB potentials are the same.
  \label{tableotherbranes}}
\end{table}

The above analysis shows that the wrapping rules in eqs.~\eqref{allwrappingrulesinonego} and \eqref{wrappingrulealpha=-4} naturally follow from the T-duality rules in eq.~\eqref{allbranesallalphasruleintro}, when applied to the potentials in Table \ref{tabletendimbranes}. We now show that from the same T-duality rules, when applied to all the other potentials of IIA and IIB that correspond to branes in lower dimensions without a ten-dimensional brane origin, one derives a set of generalised wrapping rules that allow to determine the number of all the branes in any dimension. We list in Table \ref{tableotherbranes} all such potentials both as representations of $SO(d,d)$ and as ten-dimensional mixed-symmetry potentials in the IIA and IIB theory. The rule for the values of $m,n,p,...$ giving the numbers of the mixed-symmetry indices in the third and fourth column of the table is that they take all possible values with the condition that the number of indices cannot exceed ten and that  any set of indices is larger than or equal to the one to the right. Even when all these numbers vanish, the potentials have mixed symmetry, which is a manifestation of the fact that there  are only exotic branes associated to these potentials.

\begin{table}[b!]
\begin{center}
\begin{tabular}{|c||c|c|c|c|c|c|}
\hline \rule[-1mm]{0mm}{6mm} \backslashbox{$D$}{$p$} & 0 & 1 & 2 & 3 & 4 & 5 \\
\hline \hline \rule[-1mm]{0mm}{6mm}
 7 &  & & & & & 1    \\
 \hline \rule[-1mm]{0mm}{6mm}
6 & & & & & 8 & 8 \\
 \hline \rule[-1mm]{0mm}{6mm}
5 & & & & 40& 80 &  \\
\hline \rule[-1mm]{0mm}{6mm}
4 & & & 160  & 480&  &  \\
 \hline \rule[-1mm]{0mm}{6mm}
3 & & 560 & 2240 & &  &  \\
\hline
\end{tabular}
\caption{\sl The $F_{9,3}$ family of branes, whose number is obtained starting from the 5-brane in $D=7$ by doubling times a multiplicity factor ${d \choose 3}$.
} \label{TableF93}
\end{center}
\end{table}

The T-duality rules in eq.~\eqref{allbranesallalphasruleintro} connect all the brane components of each family of potentials \cite{Lombardo:2016swq}. We can consider first the potentials $F_{9+n,3+m,m,n}$ in the first row of the table. For $m=n=0$, one gets the
potential $F_{9,3}$ which gives a 5-brane in $D=7$, corresponding to the 6-form $F_{6\, xyz, xyz}$.  By applying the T-duality rule for $\alpha=-4$, given in
eq.~\eqref{Tdualityrules-4}, one can see that $F_{6\, xyz, xyz}$ is fixed under
T-duality along $x,y,z$. This means that performing a single T-duality maps the brane in IIA to the same brane in IIB.
Compactifying to $D=6$ along the direction $w$, one gets the potentials $F_{6 \, xyz,xyz}$ (unwrapped brane) and $F_{5 \, xyzw,xyz}$ (wrapped brane).
By T-duality along $w$ the first potential goes to the component $F_{6 \, xyzw,xyzw,w,w}$ of $F_{10,4,1,1}$, while the second goes to the component $F_{5 \, xyzw,xyzw,w}$ of $F_{9,4,1}$. This means that one gets a doubling, but on top of this one should notice that in $D=6$ one  has 4 possibilities to choose the directions $x,y,z$ among the 4 compact directions, giving an additional factor of 4 for all the branes. Including this combinatorial factor, this implies that in six dimensions one gets 8 4-branes and 8 5-branes. We list in Table \ref{TableF93} the number of all branes in the $F_{9,3}$ family in any dimension. The reader can see that all these numbers result from extending to any dimension the method we have just used to obtain the branes in six dimensions. Starting from the single 5-brane in seven dimensions, one always has a doubling from the dimensional reduction, regardless of whether the brane wraps or does not wrap the circle, and on top of this one has an additional factor ${d \choose 3}$ corresponding to the choice of the second set of indices in $F_{9,3}$  among the $d$ internal indices. One gets $4 \cdot {5 \choose 3} = 40$ 3-branes and $8 \cdot  {5 \choose 3} = 80$ 4-branes in five dimensions, $8 \cdot {6 \choose 3} = 160$ 2-branes and $24 \cdot  {6 \choose 3} = 480$ 3-branes in four dimensions and, finally, $16 \cdot {7 \choose 3} = 560$ 1-branes and $64 \cdot  {7 \choose 3} = 2240$ 2-branes in three dimensions.

\begin{table}[h]
\begin{center}
\begin{tabular}{|c||c|c|}
\hline \rule[-1mm]{0mm}{6mm} \backslashbox{$D$}{$p$} & 0 & 1  \\
\hline \hline \rule[-1mm]{0mm}{6mm}
4 & & 1   \\
 \hline \rule[-1mm]{0mm}{6mm}
3 & 14 & 14  \\
\hline
\end{tabular}
\caption{\sl The $F_{8,6}$ family of branes.  These branes always double and there is an extra  multiplicity factor ${d \choose 6}$.
} \label{TableF86}
\end{center}
\end{table}

We now show how  the same rule applies to all the other branes charged with respect to the potentials in Table \ref{tableotherbranes}. There is another family of $\alpha=-4$ branes, charged under the potentials $F_{8+n,6+m,m,n}$. The first representative of this family is $F_{8,6}$, corresponding to $m=n=0$, associated to a 1-brane in four dimensions. We list in Table \ref{TableF86} the number of all branes in this family in four and three dimensions. Again, the T-duality rule of eq.~\eqref{Tdualityrules-4} implies that these branes always double, and on top of this one has to consider the additional multiplicity factor ${d \choose 6}$ corresponding to the choice of the second set of indices of $F_{8,6}$ among the $d$ internal indices. This gives $2 \cdot {7 \choose 6} =14$ 0-branes and 1-branes in three dimensions in agreement with the table.

The potentials  with $\alpha=-5$  satisfy the T-duality rule
\begin{equation}
0 \rightarrow 1,1,1,1,1 \qquad 1 \rightarrow 1,1,1,1 \qquad 1,1 \rightarrow 1,1,1 \quad . \label{Tdualityrules-5}
\end{equation}
As Table \ref{tableotherbranes} shows, there are two families of $\alpha=-5$ branes.
The first family results from the mixed-symmetry potentials $G_{10,4+n,2m+1,n}$ in IIA or  $G_{10,4+n,2m,n}$ in IIB. The number of branes is given in Table \ref{TableG1041}. In six dimensions one gets 8 5-branes from the potentials with $n=0$, {\it i.e.} $G_{10,4,1}$ and $G_{10,4,3}$ in IIA or $G_{10,4}$, $G_{10,4,2}$ and $G_{10,4,4}$ in IIB. The reader can check that by applying the rules in eq.~\eqref{Tdualityrules-5} all these potentials are mapped into each other by T-duality along any of the four internal directions. If one further compactifies to five dimensions along say the direction $x$, eq.~\eqref{Tdualityrules-5} shows that the resulting potentials are not fixed under T-duality along $x$. This implies that the branes double. Because of the choice of the second set of indices, there is an extra multiplicity factor ${d \choose 4}$. By applying this rule one gets all the numbers in Table \ref{TableG1041}.

\begin{table}[h]
\begin{center}
\begin{tabular}{|c||c|c|c|c|c|c|}
\hline \rule[-1mm]{0mm}{6mm} \backslashbox{$D$}{$p$} & 0 & 1 & 2 & 3 & 4 & 5 \\
\hline \hline \rule[-1mm]{0mm}{6mm}
6 & & & & &  & 8 \\
 \hline \rule[-1mm]{0mm}{6mm}
5 & & & & & 80 &  \\
\hline \rule[-1mm]{0mm}{6mm}
4 & & & & 480&  &  \\
 \hline \rule[-1mm]{0mm}{6mm}
3 & & & 2240 & &  &  \\
\hline
\end{tabular}
\caption{\sl The $G_{10,4,1}$ (IIA) or $G_{10,4}$ (IIB) family of branes. The branes double and there is an extra multiplicity factor ${d \choose 4}$.} \label{TableG1041}
\end{center}
\end{table}

\begin{table}[b!]
\begin{center}
\begin{tabular}{|c||c|c|c|}
\hline \rule[-1mm]{0mm}{6mm} \backslashbox{$D$}{$p$} & 0 & 1  &2\\
\hline \hline \rule[-1mm]{0mm}{6mm}
4 & & & 32  \\
 \hline \rule[-1mm]{0mm}{6mm}
3 & & 448 & 448  \\
\hline
\end{tabular}
\caption{\sl The $G_{9,6}$ (IIA) or $G_{9,6,1}$ (IIB) family of branes.} \label{TableG96}
\end{center}
\end{table}

There is another family of $\alpha=-5$ branes, coming from the potentials $G_{9+p,6+n,2m,n,p}$ in IIA or $G_{9+p,6+n,2m+1,n,p}$ in IIB. The number of branes is shown in Table \ref{TableG96}. To obtain a brane in four dimensions, one has to put $n=0$ and therefore $p=0$, giving $G_{9,6}$, $G_{9,6,2}$, $G_{9,6,4}$ and $G_{9,6,6}$ in IIA or $G_{9,6,1}$, $G_{9,6,3}$ and $G_{9,6,5}$ in IIB, giving a total of 32 2-branes. Using eq.~\eqref{Tdualityrules-5}, these potentials are all mapped into each other by T-duality along any of the six internal directions. By further compactifying to three dimensions, these branes double and there is an extra multiplicity ${d \choose 6}$ which is 7 in three dimensions, resulting in $64 \cdot 7 = 448$ 1-branes and 2-branes.

We next consider the $\alpha=-6$ branes. The T-duality rule for $\alpha=-6$ is
\begin{equation}
0 \rightarrow 1,1,1,1,1,1 \qquad 1 \rightarrow 1,1,1,1,1 \qquad 1,1 \rightarrow 1,1,1,1 \qquad 1,1,1 \rightarrow 1,1,1\quad . \label{Tdualityrules-6}
\end{equation}
From Table \ref{tableotherbranes} one reads that the family of mixed-symmetry potentials is $H_{10,6+n,2+m,m,n}$ in both IIA and IIB. The branes in this family in four and three dimensions are given in Table \ref{TableH1062}. The 240 four-dimensional 3-branes arise from the $n=0$ potentials $H_{10,6,2}$, $H_{10,6,3,1}$, $H_{10,6,4,2}$, $H_{10,6,5,3}$ and $H_{10,6,6,4}$. By compactification to three dimensions, the branes double because the wrapped branes  are not fixed under  eq.~\eqref{Tdualityrules-6}, and there is an extra multiplicity factor ${d \choose 6}$ which is 7. One therefore expects $480 \cdot 7 = 3360$
2-branes, which agrees with the table.

\begin{table}[t!]
\begin{center}
\begin{tabular}{|c||c|c|c|c|}
\hline \rule[-1mm]{0mm}{6mm} \backslashbox{$D$}{$p$} & 0 & 1  &2& 3\\
\hline \hline \rule[-1mm]{0mm}{6mm}
4 & & & & 240  \\
 \hline \rule[-1mm]{0mm}{6mm}
3 & &  &3360 &  \\
\hline
\end{tabular}
\caption{\sl The $H_{10,6,2}$ family of branes.} \label{TableH1062}
\end{center}
\end{table}

Finally, for $\alpha=-7$ the T-duality rule is
\begin{equation}
0 \rightarrow 1,1,1,1,1,1,1\ \  \quad 1 \rightarrow 1,1,1,1,1,1\ \  \quad 1,1 \rightarrow 1,1,1,1,1\ \  \quad 1,1,1 \rightarrow 1,1,1,1\  . \label{Tdualityrules-7}
\end{equation}
The mixed-symmetry potentials for the $\alpha=-7$ branes are
$I_{10,6+n,6+n,2m+1,n,n}$ in IIA and   $I_{10,6+n,6+n,2m,n,n}$ in IIB. The branes in four and three dimensions are given in Table \ref{TableI1066}. The 32 four-dimensional 3-branes result from the potentials with $n=0$, which are $I_{10,6,6,1}$, $I_{10,6,6,3}$ and $I_{10,6,6,5}$ in IIA and $I_{10,6,6}$, $I_{10,6,6,2}$, $I_{10,6,6,4}$ and $I_{10,6,6,6}$ in IIB. The 448 2-branes in three dimensions are in agreement with the doubling times a multiplicity factor ${d \choose 6}$, which is 7 in three dimensions.

\begin{table}[b!]
\begin{center}
\begin{tabular}{|c||c|c|c|c|}
\hline \rule[-1mm]{0mm}{6mm} \backslashbox{$D$}{$p$} & 0 & 1  &2& 3\\
\hline \hline \rule[-1mm]{0mm}{6mm}
4 & & & & 32  \\
 \hline \rule[-1mm]{0mm}{6mm}
3 & &  &448 &  \\
\hline
\end{tabular}
\caption{\sl The $I_{10,6,6,1}$ (IIA) or $I_{10,6,6}$ (IIB) family of branes.} \label{TableI1066}
\end{center}
\end{table}

This concludes the analysis of the branes of the maximal theory. In the next section we will discuss the branes of Type IIA and Type  IIB string theory compactified on $( T^4/\mathbb{Z}_2 )\times T^n$ and we will show that the same generalised wrapping rules apply, provided that one takes into account only the cycles that are compatible with the orbifold.

\section{Wrapping rules for IIA/IIB on $( T^4/\mathbb{Z}_2 )\times T^n$}

In \cite{Bergshoeff:2012jb,Bergshoeff:2013spa} it was shown that the wrapping rules satisfied by the branes of the maximal theory are still valid if one compactifies the Type IIA and Type IIB string theories to the six dimensional ${\cal N} = (1,1)$ and ${\cal N} = (2,0)$ theories
on the orbifold $T^4/\mathbb{Z}_2$, provided that only even cycles are taken into account.
In \cite{Pradisi:2014fqa} it was then shown that if the six-dimensional theory is further reduced on a torus, the resulting branes with $\alpha=0,-1,-2,-3$ are obtained starting from the six-dimensional ones by applying the standard wrapping rules. The aim of this section is to first review how  the branes of the six-dimensional theories result from applying the wrapping rules of the maximal theory with only even cycles taken into account, and  then move to the lower-dimensional case showing that
the analysis of  \cite{Pradisi:2014fqa} can be extended so that all the branes of the $( T^4/\mathbb{Z}_2 )\times T^n$ theory result from generalised wrapping rules, where as explained in the previous section additional multiplicity factors given by the mixed-symmetry indices of the potentials involved have to be taken into account.

We first consider the six-dimensional IIA theory compactified on $T^4/\mathbb{Z}_2$. We list in Table \ref{tablesixdimbranesIIA} the number of branes with different values of $\alpha$, together with the corresponding mixed-symmetry potentials in  ten dimensions. By denoting with $x_i$, $i=1,...,4$ the torus coordinates, the number of branes results from implementing the fact that when compactifying the mixed-symmetry potentials of the maximal theory   the total number of $x$'s must be even. The outcome of this analysis is that for $\alpha=0,-1,-2,-3$ the same number of branes can be obtained by starting from the ten-dimensional branes and applying the wrapping rules with the additional requirement that there are only even cycles. The only exception to this general rule are the 5-branes with $\alpha=-2$, which are 8 instead of 16, which is what one would naively get applying the wrapping rules. The reason of this mismatch is that the 5-branes coming from $D_{7,1}$ and $D_{9,3}$ in the maximal theory support a vector multiplet, which splits in the half-maximal theory into a vector and a hyper-multiplet, and neither of the two are allowed for $\alpha=-2$ branes in the IIA theory on  $T^4/\mathbb{Z}_2$.

The branes with $\alpha=-4$ can be obtained applying the T-duality rules in eq.~\eqref{allbranesallalphasruleintro}. We first consider the  $F_{10,2n+1,2n+1}$ family. The first representative, corresponding to $n=0$, is $F_{6 \, x_1 ... x_4 ,x_1, x_1}$, leading to four 5-branes. To remain in the same theory, we have to apply two T-dualities. If one of these T-dualities is along $x_1$ one remains in the same set of branes, while if
these T-dualities are  along $x_2$ and $x_3$, the rule in eq.~\eqref{Tdualityrules-4} gives $F_{6 \, x_1 ...x_4 , x_1 x_2 x_3, x_1 x_2 x_3}$, which is another four branes. This gives eight branes in total. Similarly, the first representative of the   $F_{9+n,3+m,m,n}$ family, for $n=m=0$,  is $F_{6 \, x_1 x_2 x_3 , x_1 x_2 x_3}$,  which corresponds to four 5-branes. Performing two T-dualities, one of which along $x_4$, one gets $F_{6 \, x_1 ...x_4 , x_1 ...x_4, x_4 ,x_4}$, which are the four brane components of the potential $F_{10,4,1,1}$. Again, one gets eight branes in total.

\begin{table}[t!]
\begin{center}
\begin{tabular}{|c||c|c|}
\hline \rule[-1mm]{0mm}{6mm} $\alpha$ & branes & 10d IIA origin\\
\hline
\hline \rule[-1mm]{0mm}{5mm}
$0$ &  1 1-brane & $B_2$\\
\hline \rule[-1mm]{0mm}{5mm}
$-1$  & 8 0-branes & $C_{2n+1}$  \\
& 8 2-branes & \\
& 8 4-branes & \\
\hline \rule[-1mm]{0mm}{5mm}
$-2$  & 1 1-brane & $D_{6+n,n}$ \\
& 24 3-branes & \\
\cline{2-3}\rule[-2mm]{0mm}{6mm}
 & 8 5-branes &  $D_{6+2n,2n}$ \\
\hline
\rule[-1mm]{0mm}{5mm}
$-3$ & 32 4-branes & $E_{8+n,2m+1,n}$ \\
\hline \rule[-1mm]{0mm}{5mm}
$-4$ & 8 5-branes & $F_{10,2n+1,2n+1}$ \\
\cline{2-3} \rule[-1mm]{0mm}{5mm}
 & 8 5-branes & $F_{9+n,3+m,m,n}$\\
\hline
\end{tabular}
\end{center}
  \caption{\sl  The  branes of the  IIA theory compactified on $T^4/\mathbb{Z}_2$ and their corresponding ten-dimensional origin in terms of mixed-symmetry potentials.
  \label{tablesixdimbranesIIA}}
\end{table}

In the case of the IIB theory compactified on $T^4/\mathbb{Z}_2$, we list the number of branes and the corresponding IIB mixed-symmetry potentials in  Table \ref{tablesixdimbranesIIB}. Again, the number of branes results from implementing in the compactification of the mixed-symmetry potentials of the maximal theory the fact that the total number of $x$'s must be even. For $\alpha=0,...,-4$,
the resulting numbers can be obtained starting  from the ten-dimensional branes and applying the wrapping rules with the additional requirement that there are only even cycles. Exactly as in the IIA theory, the $\alpha=-2$ 5-branes are an exception to this rule, because there are only 8 branes instead of 16. Again, the reason is that the 5-branes coming from $D_{7,1}$ and $D_{9,3}$ in the maximal theory support tensor multiplets, which split into tensor and hyper-multiplets in the ${\cal N} = (2,0)$ theory, which are both not allowed. Actually, exactly for the same reason the $F_{9+n,3+m,m,n}$ family of mixed-symmetry potentials of the maximal theory is projected out in the $(2,0)$ theory. Finally, in this case there are also eight branes coming from the $ G_{10, 4+n,2m, n}$ family. The first representative of  this family is $G_{6 \, x_1 ..x_4 , x_1 ...x_4}$, corresponding to $m=n=0$, and by applying two T-dualities (six possibilities) and four T-dualities (one possibility) and using eq.~\eqref{Tdualityrules-5} one gets eight branes in total, corresponding to all the brane components in the family.

\begin{table}[h]
\begin{center}
\begin{tabular}{|c||c|c|}
\hline \rule[-1mm]{0mm}{6mm} $\alpha$ & branes  & 10d IIB origin\\
\hline
\hline \rule[-1mm]{0mm}{5mm}
$0$ &  1 1-brane & $B_2$\\
\hline \rule[-1mm]{0mm}{5mm}
$-1$ & 8 1-branes  & $C_{2n}$  \\
& 8 3-branes & \\
& 8 5-branes & \\
\hline \rule[-1mm]{0mm}{5mm}
$-2$ & 1 1-brane & $D_{6+n,n}$ \\
& 24 3-branes & \\
\cline{2-3}
\rule[-2mm]{0mm}{6mm} & 8 5-branes &  $D_{6+2n,2n}$ \\
\hline \rule[-1mm]{0mm}{5mm}
$-3$ & 8 3-branes & $E_{8+n,2m,n}$ \\
& 48 5-branes & \\
\hline \rule[-2mm]{0mm}{6mm}
$-4$ & 8 5-branes & $F_{10,2n,2n}$ \\
\hline \rule[-2mm]{0mm}{6mm}
$-5$ & 8 5-branes & $G_{10,4,2m}$ \\
\hline
\end{tabular}
\end{center}
  \caption{\sl The  branes of the  IIB theory compactified on $T^4/\mathbb{Z}_2$ and their corresponding ten-dimensional origin in terms of mixed-symmetry potentials.
  \label{tablesixdimbranesIIB}}
\end{table}

We now move on to discuss the branes of the lower-dimensional theories. The number of branes of these theories for different values of $\alpha$ have been obtained  in \cite{Pradisi:2014fqa} by identifying the dilaton scaling inside the non-perturbative symmetry of the half-maximal theory. Once this symmetry is decomposed with respect to the perturbative one, the number of branes is given as in the maximal theory by counting the number of long weights of the representation.\footnote{In the case of non-split groups, one has actually to impose the additional requirement that the long weight is real, where the reality properties are defined by the Tits-Satake procedure \cite{Bergshoeff:2014lxa}.} What was then shown in \cite{Pradisi:2014fqa} is that for the branes with $\alpha=0,-1,-2,-3$ the number of branes are those that one obtains by applying the wrapping rules starting from six dimensions. Here we want to refine this analysis and derive from the generalised wrapping rules the number of  branes with more-negative values of $\alpha$ in any dimension. We refer to tables 3, 4 and 5 of \cite{Pradisi:2014fqa} for the numbers of branes with different values of $\alpha$ in dimension 5, 4 and 3 respectively.

We start by considering the $F_{10,2n+1,2n+1}$ (IIA) or $F_{10,2n,2n}$ (IIB) family of mixed-symmetry potentials. We denote as before with $x$ the orbifold directions, and  we denote with $y$ the torus coordinates. By computing all allowed brane components in any dimension, one arrives at the numbers listed in Table \ref{TableF10onK3}. It is straightforward to see that the wrapping rules apply exactly as in the maximal case, and the branes always double.

\begin{table}[h]
\begin{center}
\begin{tabular}{|c||c|c|c|c|c|c|}
\hline \rule[-1mm]{0mm}{6mm} \backslashbox{$D$}{$p$} & 0 & 1 & 2 & 3 & 4 & 5 \\
\hline \hline \rule[-1mm]{0mm}{6mm}
6A/6B & & & & &  & 8/8 \\
 \hline \rule[-1mm]{0mm}{6mm}
5 & & & & & 16 &  \\
\hline \rule[-1mm]{0mm}{6mm}
4 & & & & 32 &  &  \\
 \hline \rule[-1mm]{0mm}{6mm}
3 & & & 64 & &  &  \\
\hline
\end{tabular}
\caption{\sl The $F_{10,1,1}$ (IIA) or $F_{10}$ (IIB) family of branes for the $( T^4/\mathbb{Z}_2 )\times T^n$ theories. Here and in the next tables we denote with 6A and 6B the ${\cal N} =(1,1)$ and ${\cal N} =(2,0)$ theories, respectively.} \label{TableF10onK3}
\end{center}
\end{table}

We then move to the  $F_{9+n,3+m,m,n}$ family. To derive the number of branes in this case, one should remember, as we already mentioned above,  that in  the ${\cal N} =(2,0)$ theory in six dimensions the components $F_{6 \, x_1 x_2 x_3, x_1 x_2 x_3}$ and $F_{6 \, x_1 ...x_4 , x_1 ...x_4 , x_1 , x_1}$ are projected out, and therefore they remain projected out after dimensional reduction. This means for instance that if one compactifies on $y$ and wants to derive the number of branes in the IIB picture, the components $F_{5 \, x_1 x_2 x_3 y, x_1 x_2 x_3}$ and $F_{5 \, x_1 ...x_4 y, x_1 ...x_4 , x_1 , x_1}$ should not be included, and equivalently, T-dualising along $y$,  in the IIA picture the components $F_{5 \, x_1 x_2 x_3 y, x_1 x_2 x_3 y,y}$ and $F_{5 \, x_1 ...x_4 y, x_1 ...x_4 y, x_1 y, x_1}$ should be ignored. The final result is that one ends up with the number of branes given in Table \ref{TableF93onK3}.

\begin{table}[b!]
\begin{center}
\begin{tabular}{|c||c|c|c|c|c|c|}
\hline \rule[-1mm]{0mm}{6mm} \backslashbox{$D$}{$p$} & 0 & 1 & 2 & 3 & 4 & 5 \\
\hline \hline \rule[-1mm]{0mm}{6mm}
6A/6B & & & & &  & 8/0 \\
 \hline \rule[-1mm]{0mm}{6mm}
5 & & & & 24 & 8 &  \\
\hline \rule[-1mm]{0mm}{6mm}
4 & & &96 & 208 &  &  \\
 \hline \rule[-1mm]{0mm}{6mm}
3 & & 304 & 1184 & &  &  \\
\hline
\end{tabular}
\caption{\sl The $F_{9,3}$ family of branes  for the $( T^4/\mathbb{Z}_2 )\times T^n$ theories.} \label{TableF93onK3}
\end{center}
\end{table}

We want to determine these numbers from the generalised wrapping rules. We start from six dimensions, where as we have already mentioned we have
\begin{equation}
D=6: \qquad F_{6 \, x_1 x_2 x_3, x_1 x_2 x_3} \qquad \rightarrow \qquad 4 \times 2 =8 \ \ {\rm (only \ IIA)} \quad ,
\end{equation}
where 4 gives the multiplicity and 2 the doubling corresponding to performing two T-dualities. By dimensional reduction, in lower dimensions we get
\begin{eqnarray}
& & D=5: \qquad F_{5 \, x_1 x_2 x_3 y, x_1 x_2 x_3} \qquad \qquad \rightarrow \qquad 4 \times 2 =8 \quad ,\nonumber \\
& & D=4: \qquad F_{4 \, x_1 x_2 x_3 y_1 y_2, x_1 x_2 x_3} \qquad \quad \rightarrow \qquad 4 \times 2 \times 2 =16 \quad , \nonumber \\
& & D=3: \qquad F_{3 \, x_1 x_2 x_3 y_1 y_2 y_3, x_1 x_2 x_3} \ \qquad \rightarrow \qquad 4 \times 2 \times 4=32  \quad ,\label{firstsetF93orbifold}
\end{eqnarray}
where the extra factor of 2 and of 4 in four  and three dimensions is the doubling due to the T-dualities in the $y$ directions. For these components, starting from six dimensions, the lower-dimensional numbers simply result from the wrapping rules.

Apart from the components of $F_{9,3}$ listed in eq.~\eqref{firstsetF93orbifold}, that arise from the dimensional reduction of the six-dimensional one, there are additional components that can arise in lower dimensions due to the index structure. In particular, in five dimensions one can have $F_{4 \, x_1 ...x_4 y, x_1 x_2 y}$, which is allowed because there is an even number of $x$ indices, but does not arise from six dimensions. One can determine the multiplicity of the family of branes that result in any dimension as usual using the T-duality rules in eq.~\eqref{Tdualityrules-4}. The result is
 \begin{eqnarray}
& & D=5: \quad  F_{4 \, x_1 x_2 x_3 x_4 y, x_1 x_2 y} \qquad \quad \  \rightarrow \qquad 6 \times 2 \times 2 =24 \nonumber \\
& & D=4: \quad  F_{3 \, x_1 x_2 x_3 x_4 y_1 y_2, x_1 x_2 y_1} \qquad \rightarrow \qquad 24 \times 2 \times 2 =96  \nonumber \\
& & \qquad \quad \quad \ \,  F_{4 \, x_1 x_2 x_3 x_4  y_1 , x_1 x_2 y_1} \ \ \qquad \rightarrow \qquad 24 \times 2 \times 2 =96 \nonumber \\
&  & D=3: \quad  F_{2 \, x_1 x_2 x_3 x_4 y_1 y_2 y_3, x_1 x_2 y_1}   \quad \rightarrow \qquad 24 \times 4 \times 3=288 \nonumber \\
& & \qquad \quad \quad \ \,  F_{3 \, x_1 x_2 x_3 x_4 y_1 y_2, x_1 x_2 y_1} \qquad \rightarrow \qquad 24 \times 4 \times 6=576 \label{secondsetF93orbifold}
 \quad ,
\end{eqnarray}
where in five dimensions the factors 2 arise from T-dualities along $x$, while in four and three dimensions the second factor arise from T-dualities along $y$ directions and the third factor from the choice of $y$ indices. The reader can check that all the numbers in four and three dimensions in eq.~\eqref{secondsetF93orbifold} are given by applying the wrapping rules on the 24 3-branes in 5D supplemented by a factor $6-D$.

In four dimensions there is an additional component $F_{4 \, x_1 x_2 x_3 y_1 y_2 , x_1 y_1 y_2}$ that cannot arise from higher dimensions. The number of corresponding 3-branes in the family is $12\times 8$, where 12 is due to the choices of $x$'s and 8 from all the possible non-trivial T-dualities. One can also determine the branes in three dimensions, and the final result is
 \begin{eqnarray}
& & D=4: \quad  F_{4 \, x_1 x_2 x_3 y_1 y_2, x_1  y_1 y_2} \qquad \rightarrow \qquad 12 \times 8 =96  \nonumber \\
&  & D=3: \quad  F_{3 \, x_1 x_2 x_3 y_1 y_2 y_3, x_1  y_1 y_2}  \ \quad \rightarrow \qquad 96 \times 2 \times 3=576  \label{thirdsetF93orbifold}
 \quad ,
\end{eqnarray}
and again the number of 2-branes in three dimensions is given by the wrapping rules times an extra multiplicity factor, which is ${6-D \choose 2}$ in this case. Finally, in three dimensions there is the extra possibility
\begin{equation}
D=3: \qquad F_{2 \, x_1 x_2 x_3 x_4 y_1 y_2 y_3, y_1 y_2 y_3} \qquad \rightarrow \qquad 2^4 =16  \quad ,
\end{equation}
giving in total 16 branes because one can perform T-duality trasformations along all $x$ directions, while the component is fixed under T-dualities along the $y$ directions.

This concludes the analysis of the branes in the $F_{9,3}$ family. One can check that the numbers we have derived using the generalised wrapping rules reproduce Table \ref{TableF93onK3}. For instance, from eqs.~\eqref{firstsetF93orbifold}, \eqref{secondsetF93orbifold} and \eqref{thirdsetF93orbifold} one gets $16+96+96=208$ 3-branes in four dimensions and $32+576+576 = 1184$ 2-branes in three dimensions, which coincides with Table \ref{TableF93onK3}. Similarly, all the other numbers can easily be checked.

The last family of $\alpha=-4$ potentials is the $F_{8+n,6+m,m,n}$ family, whose first representative is $F_{8,6}$ which is relevant in four dimensions and below. By computing all possible brane components in the family one gets the numbers that are given in Table \ref{TableF86onK3}.
In four dimensions the only brane component is $F_{2 \, x_1 x_2 x_3 x_4 y_1 y_2 , x_1 x_2 x_3 x_4 y_1 y_2 }$. This is fixed under all T-dualities, and therefore has multiplicity 1. By dimensional reduction one gets
 \begin{eqnarray}
& & D=4: \quad  F_{2 \, x_1 x_2 x_3 x_4 y_1 y_2, x_1  x_2 x_3 x_4 y_1 y_2} \qquad \rightarrow \qquad 1  \nonumber \\
&  & D=3: \quad  F_{1 \, x_1 x_2 x_3 x_4 y_1 y_2 y_3, x_1 x_2 x_3 x_4  y_1 y_2}   \quad \ \rightarrow \qquad 2 \times 3 =6 \nonumber \\
& & \qquad \quad \ \ \quad  F_{2 \, x_1 x_2 x_3 x_4 y_1 y_2, x_1  x_2 x_3 x_4 y_1 y_2} \qquad  \rightarrow \qquad 2 \times 3 =6   \label{firstsetF86orbifold}
 \quad ,
\end{eqnarray}
and again in three dimensions one has the doubling times a factor 3 due to the $y$ indices. Finally, in three dimensions one has the additional component
\begin{equation}
D=3: \qquad F_{2 \, x_1 x_2 x_3 y_1 y_2 y_3, x_1 x_2 x_3 y_1 y_2 y_3} \qquad \rightarrow \qquad 2 \times 4 = 8  \quad .\label{secondsetF86orbifold}
\end{equation}
It is easy to check that the branes in eqs.~\eqref{firstsetF86orbifold} and \eqref{secondsetF86orbifold} reproduce Table \ref{TableF86onK3}.

\begin{table}[h]
\begin{center}
\begin{tabular}{|c||c|c|}
\hline \rule[-1mm]{0mm}{6mm} \backslashbox{$D$}{$p$} & 0 & 1 \\
\hline \hline \rule[-1mm]{0mm}{6mm}
4 & & 1  \\
 \hline \rule[-1mm]{0mm}{6mm}
3 & 6 & 14   \\
\hline
\end{tabular}
\caption{\sl The $F_{8,6}$ family of branes for the $( T^4/\mathbb{Z}_2 )\times T^n$ theories.} \label{TableF86onK3}
\end{center}
\end{table}

This concludes the analysis of the $\alpha=-4$ branes in the orbifold theory. By putting together the numbers in Tables \ref{TableF10onK3}, \ref{TableF93onK3} and \ref{TableF86onK3}, it is straightforward to check that the overall numbers exactly reproduce tables 3, 4 and 5 of \cite{Pradisi:2014fqa}.

One can show that using the generalised wrapping rules one can  also determine the number of branes with more negative values of $\alpha$ in the orbifold theory. As an example we consider the $\alpha=-5$ case. We give in Table \ref{TableGonK3} the number of such branes as derived in \cite{Pradisi:2014fqa}. The mixed-symmetry potentials that contribute are the $ G_{10, 4+n,2m+1, n}$ (IIA) or $ G_{10, 4+n,2m, n}$ (IIB) family and the $ G_{9+p, 6+n,2m, n,p}$ (IIA) or  $ G_{9+p, 6+n,2m+1, n,p}$ (IIB) family. We determine in any dimension each component that is a representative of a T-duality family in the orbifold theory and we determine the number of branes in the family by simply looking at how the representative transforms under the T-duality rules given in eq.~\eqref{Tdualityrules-5}.  We take the IIB representatives for the first family and the IIA representative for the second, but one can always map this to the other theory by a single T-duality in a $y$ direction.
Starting from  $G_{6 \, x_1 ... x_4, x_1 ... x_4}$  in six dimensions  we get
\begin{eqnarray}
& & D=6: \qquad G_{6 \, x_1 ... x_4, x_1 ... x_4} \qquad \quad \quad \ \rightarrow \qquad 8 \ \ {\rm (only \ IIB)} \nonumber \\
& & D=5: \qquad G_{5 \, x_1 ... x_4  y, x_1 ... x_4} \qquad \qquad \rightarrow \qquad 8 \nonumber \\
& & D=4: \qquad G_{4 \, x_1 ...x_4 y_1 y_2, x_1 ...x_4 } \qquad \quad \rightarrow \qquad 16  \nonumber \\
& & D=3: \qquad G_{3 \, x_1 ...x_4 y_1 y_2 y_3, x_1 ...x_4 } \ \qquad \rightarrow \qquad 32  \quad , \label{firstsetGorbifold}
\end{eqnarray}
and in this case the wrapping rule applies.
The family $G_{5 \, x_1 ...x_4 y, x_1 x_2 x_3 y, x_1 y}$ gives
\begin{eqnarray}
& & D=5: \qquad G_{5 \, x_1 ... x_4  y, x_1 x_2 x_3 y, x_1 y} \qquad \qquad \ \ \rightarrow \qquad 32 \nonumber \\
& & D=4: \qquad G_{4 \, x_1 ...x_4 y_1 y_2, x_1 x_2 x_3 y_1 , x_1 y_1} \qquad \quad \rightarrow \qquad 64 \times 2 = 128 \nonumber \\
& & D=3: \qquad G_{3 \, x_1 ...x_4 y_1 y_2 y_3, x_1 x_2 x_3 y_1 , x_1 y_1} \ \qquad \rightarrow \qquad 128 \times 3 = 384 \quad , \label{secondsetGorbifold}
\end{eqnarray}
and the doubling is corrected by a multiplicity factor $6-D$ due to the $y$ index.
One also has
\begin{eqnarray}
& & D=4: \qquad G_{4 \, x_1 ...x_4 y_1 y_2, x_1 x_2 y_1 y_2 } \qquad \quad \rightarrow \qquad 96 \nonumber \\
& & D=3: \qquad G_{3 \, x_1 ...x_4 y_1 y_2 y_3, x_1 x_2 y_1 y_2} \ \qquad \rightarrow \qquad 192 \times 3 =  576 \quad , \label{thirdsetGorbifold}
\end{eqnarray}
where the multiplicity is ${6-D \choose 2}$. Finally, the $ G_{10, 4+n,2m, n}$ also produces in three dimensions
\begin{equation}
D=3: \qquad G_{3 \, x_1 ...x_4 y_1 y_2 y_3, x_1 y_1 y_2 y_3 , x_1 y_1} \qquad \rightarrow \qquad 128  \quad .\label{fourthsetF86orbifold}
\end{equation}
We then move to the $ G_{9+p, 6+n,2m, n,p}$ IIA family. The highest dimensional representative is $G_{3 \, x_1 ...x_4 y_1 y_2, x_1 ...x_4 y_1 y_2}$, giving
\begin{eqnarray}
& & D=4: \qquad G_{3 \, x_1 ...x_4 y_1 y_2  , x_1 ...x_4 y_1 y_2 } \qquad \quad \rightarrow \qquad 16 \nonumber \\
& & D=3: \qquad G_{2 \, x_1 ...x_4 y_1 y_2 y_3, x_1 ...x_4  y_1 y_2} \ \qquad \rightarrow \qquad 32 \times 3 = 96 \nonumber \\
& & \qquad \quad \quad \ \ \quad G_{3 \, x_1 ...x_4 y_1 y_2  , x_1 ...x_4 y_1 y_2 } \qquad \quad \rightarrow \qquad 32 \times 3 =96
\quad , \label{fifthsetGorbifold}
\end{eqnarray}
satisfying the wrapping rules with an extra factor ${6-D \choose 2}$. In three dimensions one can also have
\begin{equation}
D=3: \qquad G_{2 \, x_1 ...x_4 y_1 y_2 y_3, x_1 x_2 x_3 y_1 y_2 y_3 , x_1 y_1} \qquad \rightarrow \qquad 128  \label{7thsetF86orbifold}
\end{equation}
and
\begin{equation}
D=3: \qquad G_{3 \, x_1 x_2 x_3  y_1 y_2 y_3, x_1 x_2 x_3 y_1 y_2 y_3 } \qquad \rightarrow \qquad 128 \quad . \label{7thsetF86orbifold}
\end{equation}
The reader can check that summing all the $\alpha=-5$ branes one recovers Table \ref{TableGonK3}. We leave it as an exercise to show that the generalised wrapping rules are valid also for the branes with lower values of $\alpha$.

\begin{table}[h]
\begin{center}
\begin{tabular}{|c||c|c|c|c|c|c|}
\hline \rule[-1mm]{0mm}{6mm} \backslashbox{$D$}{$p$} & 0 & 1 & 2 & 3 & 4 & 5 \\
\hline \hline \rule[-1mm]{0mm}{6mm}
6A/6B & & & & &  & 0/8\\
 \hline \rule[-1mm]{0mm}{6mm}
5 & & & &  & 40 &  \\
\hline \rule[-1mm]{0mm}{6mm}
4 & & &16 & 240 &  &  \\
 \hline \rule[-1mm]{0mm}{6mm}
3 & & 224 & 1344 & &  &  \\
\hline
\end{tabular}
\caption{\sl The $\alpha=-5$ branes  in the $( T^4/\mathbb{Z}_2 )\times T^n$ theories.} \label{TableGonK3}
\end{center}
\end{table}

\section{Conclusions and Outlook}

The main message of this paper is that one can formulate universal wrapping rules for the branes of Type IIA and IIB string theory
that reproduce the number of branes in different $D<10$ dimensions. For the case of maximal supersymmetry we presented the basic wrapping rules in
eqs.~\eqref{allwrappingrulesinonego} and \eqref{wrappingrulealpha=-4}. These rules are valid for all branes that, by T-duality, are related to a brane which can be obtained from a 10D brane via compactification. We extended these rules to a set of generalized wrapping rules that are even valid for families of branes that are not related, via duality, to any brane with a 10D brane origin. The generalization consists of the fact that the number of branes produced by the basic wrapping rules \eqref{allwrappingrulesinonego} and \eqref{wrappingrulealpha=-4} must be multiplied by an additional combinatorial factor that is determined by the mixed-index structure of the 10D potential that gives rise to the highest-dimensional brane in the family of branes under consideration.

We also considered a case with half-maximal supersymmetry corresponding to the compactification of Type IIA or Type IIB string theory over $(T^4/{\mathbb{Z}_2}) \times T^n$. We found that the same generalized wrapping rules apply to this case too but that not all branes in $D<6$ dimensions can be obtained by compactification of the 6D branes that arise after the compactification over $T^4/{\mathbb{Z}_2}$. Additional branes, generating new families of branes, pop up in  $D<6$ dimensions. We found that  the number of branes in each such a new family is determined by the same generalized wrapping rules we constructed before. These new branes can be found by systematically investigating the components of the 10D mixed-symmetry potentials that are allowed by the orbifold and torus reduction. An additional subtlety that occurred in our analysis was the fact that a few seemingly allowed mixed-symmetry components were projected out due to the fact that they corresponded to a world-volume theory that was not allowed in the corresponding string theory.

Our analysis of the half-maximal case relies on the fact that we have a natural way of implementing the action of $\mathbb{Z}_2$ on the mixed-symmetry potentials.
It would be interesting to see how the procedure can be extended to the other possible orbifold limits $T^4/{\mathbb{Z}_N}$ of K3, with $N=3,4,6$.  The generalisation to arbitrary  K3's is a non-trivial  challenge  due to the fact that our method relies on T-duality, which is not well-defined on a generic K3 manifold. On the other hand, the group-theory analysis of \cite{Pradisi:2014fqa} shows that the number of 1/2-BPS single-brane states does not depend on the particular choice of K3, which implies that the wrapping rules still apply, although their interpretation in terms of mixed-symmetry potentials and T-duality rules is not clear.

The remarkable thing about the wrapping rules we formulated in this paper is that they seem to be universal. Independent of how complicated the T-duality representations are, especially in the case of non-maximal supersymmetry, at the end of the day the BPS $p$-branes that hide within these complicated representations satisfy the same simple set of generalized wrapping rules. This simplicity suggests that the wrapping rules have something to say about the stringy geometry that is probed by these BPS $p$-branes.

It would be interesting to compare our results with those of Double Field Theory (DFT) \cite{Siegel:1993th,Hull:2009mi,Hohm:2010jy} where the T-duality is made manifest by doubling the spacetime coordinates.
Mixed-symmetry potentials do enter also DFT as soon as one tries to dualise the NS-NS 2-form since by T-duality this 2-form is related to the metric \cite{Bergshoeff:2016ncb}.
Hence, to dualise the 2-form into a 6-form in a  T-duality covariant way one should also dualise the metric which is a notoriously difficult issue. At the linearised level \cite{Hull:2000zn,West:2001as}
this dualisation leads to a mixed-symmetry potential of the type $D_{7,1}$ that couples to the 10D Kaluza-Klein monopole. Unfortunately, we do not know how to extend this dualisation procedure to the non-linear case \cite{Bekaert:2002uh,Bekaert:2004dz}. This is the main stumbling block of our present approach.  Being able to define mixed-symmetry potentials at the non-linear level will without doubt lead to crucial insights into what the true nature of the elusive stringy  geometry is.

\vskip .7cm

\section*{Acknowledgments}

This work started at the GGI workshop `Supergravity: what next?' in 2016. We thank the Galileo Galilei Institute for Theoretical Physics for the hospitality and the INFN for partial support during the initiation of this work. We wish to thank the Universidad Aut\'onoma de Madrid (UAM), the  Rudjer Bo\v skovi\' c Institute in Zagreb (Croatia) and the Centro de Sciencias de Benasque Pedro Pascual (Spain) for hospitality and providing an inspiring environment.
One of us (F.R.) wishes to thank the Van Swinderen Institute of the University of Groningen for hospitality.

\end{document}